# A Stock Prediction Model Based on DCNN

Qiao Zhou    Ningning Liu※

(School of Information Technology and Management, University of International Business and

Economics, BJ, 100029, China)

**Abstract:** Stock market price prodiction has always been a challenging issue, due to its volatile nature. It can be affected by many factors such as national policies, companies financial reports, industry trends, and investors sentiment etc. In this paper, we present a prediction model based on deep Convolutional Neural Networks(CNN) and the candle charts, the continuous time stock information is processed. According to different information richness, prediction time interval and classification method, the original data is divided into multiple categories as the training set of CNN. In addition, the Convolutional Neural Network is used to predict the stock market and analyse the difference in accuracy under different classification methods.

The results show that the method has the best performance when the forecast time interval is 20 days. Moreover, the Moving Average Convergence Divergence(MACD) and three kinds of moving average are added as input. This method can accurately predict the stock trend of the US NDAQ exchange for 92.2%. Meanwhile, this paper evaluates three common classification methods and compares their impact on prediction accuracy.

## 1. Introduction

The stock market movement is in correlation with macroeconomy. The correct prediction of the stock market is conducive to the country's timely adjustment of macroeconomic policies and the maintenance of the stable development of the market and society. Therefore, the correct prediction of the trend of the stock market has become a hot issue in the field. However, there are many factors affecting the stock price, such as national policies, corporate financial statements, industry performance and other factors, which make the stock trend prediction become a very challenging problem.

According to Fama's Efficient Market Hypothesis(EMH) [1], under the condition of complete market information, investors still can't obtain returns beyond the average market profit through fundamental analysis and technical analysis, because of the high noise in the financial time series. In the traditional statistical methods, due to the problem of the amount of data, the complete financial time series must be preprocessed, and then the processing results are input into the model, but this processing result will directly destroy the authenticity of the data. Chulho Jung etc [2] used Vector Autoregression (VaR) model, Error Correction Model (ECM) and Kalman Filter Model (KFM) to predict the long-term changes of the UK stock market. However, due to the frequent fluctuations of the stock market, it is impossible to judge the short-term changes of the stock market, and draw the conclusion that the short-term stock market can not be predicted under the current technical means. Wei Weixian and Zhou Xiaoming [3] used GARCH model, QGARCH model and GJR model to accurately predict the fluctuation of China's stock market, but they could not grasp the trend of the stock market. Li Haitao [4] used Markov method to analyze the closing price of Shanghai Stock Northeast Expressway and predict its stock price change. Although this method has a certain success rate, it can not accurately predict the daily trend of stock price. To sum up, due to the complexity of financial time series, it is difficult to predict the trend of short-term stocks by using statistical methods [5].

With the development of information technology and the improvement of computing capability, deep learning algorithm has been widely used in medicine [6], agronomy [7], engineering [8] and other fields, which provides a new possibility for short-term stock market prediction. Luca Di Persio and others[9] used Multi-layer Lerceptron (MLP), Convolutional Neural Network (CNN), Long Short Term Memory (LSTM) and Recurrent Neural Network (RNN) to predict the price movement of S&P500 index. The results show that the CNN model is the best. Yang Jiao [10] Based on deep network, through the comparison of standard cross validation, sequential verification and single validation methods, he found that the prediction accuracy of S&P 500 index can be greatly improved by using recent information, such as closed European and Asian indexes. Gozde sismanoglu and others[11] used IBM stock information from 1968 to 2018 and used MLP and CNN algorithm to predict the information in the database. The results show that the method has good accuracy for specific stocks, and can accurately predict the rise and fall of stocks on the next day. Therefore, it can be said that it is feasible to predict the short-term stock market by using CNN method.

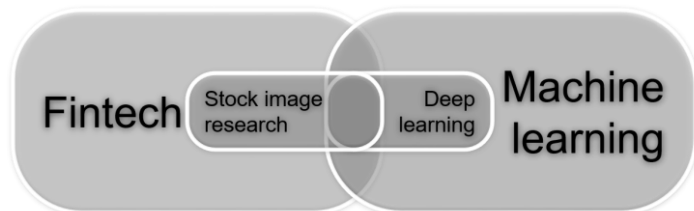

**Figure 1.** The cross field between machine learning and fintech

Based on the above research, this paper uses Convolutional Neural Network in deep learning algorithm to predict the trend of NASDAQ stock market, and integrates the stock chart, abandons the traditional data input, in order to seek a new breakthrough. Four kinds of stock charts with different information richness are processed from financial time series data, and the image data set is used as the input of modern machine learning algorithm CNN. The accuracy of forecasting stock trend in the future 1 day, 20 days, 30 days and 90 days is compared, and the horizontal comparison of three different classification training methods and the influence on the training accuracy when put into the CNN model are compared.

## 2. Method

### 2.1 Stock Images

As a financial chart, stock charts can be used to describe stock price movements in a given time period. The stock chart was invented by a Japanese rice trader called Munehisa Hooma[16], so it is also known as the Japanese candlestick chart. Each candlestick usually displays one day of transaction data, so a monthly chart can convert 20 trading days of data into 20 candlesticks. Figure 2 shows the information contained in the candlestick chart. The candlestick usually consists of three parts, which are the upper hatching, the lower hatching and the solid, it also contains four important components of the trading day information, namely the opening price, closing price, low price and high price. If the opening price is higher than the closing price, it means that the stock price is rising. This is called a bullish candlestick chart, and the main body is filled with red; otherwise, it is called a bearish candlestick chart and is filled with green. The upper and lower hatched lines represent the high and low price ranges within the specified time period, respectively, so not all candlesticks have hatched lines. However, not all candlesticks have shadows.

Stock charts are the visualisation method to assist stock trading decisions. By using the stock chart, traders can understand the stock market directions more easily [17]. This article mainly uses the visual image, that is, the stock image to participate in the DCNN training process. Through the characteristics of the stock image and the advantages of CNN vision and image processing, to achieve the purpose of accurate prediction of stock trends.

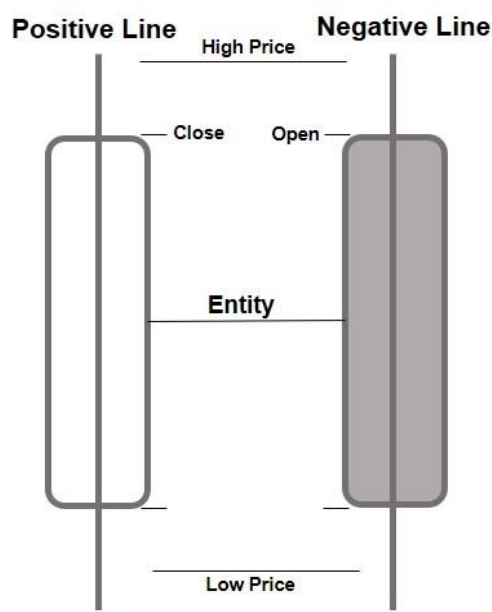

**Figure 2.** Candlestick chart diagram

### 2.2 DCNN Model

In this study, CNN-based Deep Learning Networks(DLN) are used to predict stock market trends. CNN are a class of deep learning algorithms that are currently used for image processing tasks, such as target detection [18], image classification [19], image segmentation [20], etc., demonstrating the recognition accuracy far beyond traditional methods. CNN requires that the collected data information is time continuous, that is, the value of a certain pixel is related to its neighboring pixels, and CNN is easier to extract texture information and edge information than the fully connected layer, thereby improving the prediction effect.

CNN is a feed-forward artificial neural network with multiple layers, as shown in Figure 3. The hidden layer of CNN is usually composed of pooling layer, convolution layer and full connection layer. The convolution layer is responsible for reading small pieces of data and using the kernel to read inputs such as two-dimensional images or one-dimensional signals, and scan the entire input field. The pooling layer uses feature projection, and finally the output of the pooling layer is sent to one or more fully connected layers, which will interpret what has been read and map this internal representation to class values. It is similar to a general neural network (NN) composed of a group of neurons with learning weights and bias capabilities. The difference is that the convolution layer uses convolution operations to input and then transmits the results to the next layer. This operation allows for more efficient forward functionality with fewer parameters.

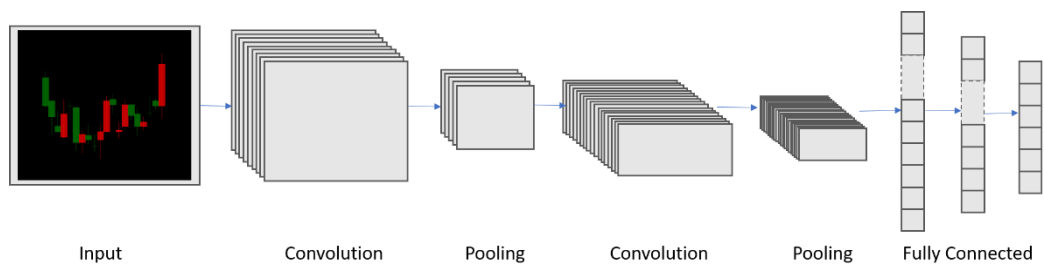

Figure 3. Schematic diagram of CNN network structure

In order to make the prediction results more accurate, this article trains the CNN model. The results of this paper also prove that CNN is very effective in addressing the problems of computer vision and image processing. This section builds a CNN-based network model for the pictures of the contained financial time series data. The network model is composed of four 2d convolutional layers, four 2d maximum pooling layers, and three output layers, as shown in Table 1.

Table 1. CNN structure for stock trend prediction

| Input |
|---|
| Conv2D-32 ReLU |
| Max-poolling |
| Conv2D-48 ReLU |
| Max-poolling |
| Dropout |
| Conv2D-64 ReLU |
| Max-poolling |
| Conv2D-96 ReLU |
| Max-poolling |
| Dropout |
| Flatten |
| Dense-256 |
| Dropout |
| Dense-2 |

## 3. Test Process

### 3.1 Sample Selection and Experiment Settings

Obtain sufficient data is vital to ensure the CNN model to predict successfully. This article is based on the Application Program Interface (API) service of Yahoo!, which collects the transaction data of 100 stocks of NDAQ exchange. To be noted that the trading date are not always in sequence from Monday to Friday. Therefore, the data crawler had made selection on the data. The date of the collected data is shown in Table 2.

Table 2. Data type and division

| Stock Data | Training Data | | Testing Data | |
|---|---|---|---|---|
| | Start | End | Start | End |
| NDAQ | 2014/12/31 | 2018/12/31 | 2019/1/1 | 2019/12/31 |

### 3.2 Data Classification

In order to analyze the changes of different data classification methods on the final accuracy, this article adopts three methods to to process the collected data, they are: random classification, automatic classification and time division classification.

### 3.2.1 Random Classification

This article uses the train_test_split function in the sklearn library for random classification. The ratio of the test set and training set can be determined according to the test_size set in advance. This paper randomly selects a certain share of candlesticks as a test set according to a set ratio, and randomly divides the data set. This function can set random_state to be an integer or None to control whether the data generated each time is always random or only once to retain the results. The ratio of random classification in this paper is set to 0.2.

3.2.2 Automatic Division

Autonomous division is achieved by intercepting the data set, the input data is intercepted into a training set and a testing set at a certain ratio, that is, no stocks are distinguished, all data are treated equally, and the data is divided in a one-size-fits-all manner. The final test set may lack some stocks data. This method has high applicability to single stocks or similar stocks. In this paper, the ratio of autonomous classification is set to 0.2.The ratio of the training set is 0.8, and the rest is the testing set.

3.2.3 Time Division

Time division, as the name implies, uses the time characteristics of financial time series, according to the trading time of each stock, set the date in advance, and set the data after the test date as the prediction set, the final test set contains part of the data for each stock . In this article, a total of 5 years of trading data of NDAQ stocks were collected. In order to compare the first two classification methods, the data between 2019/1/1 and 2019/12/31 was extracted as the test set.

3.3  Data Preprocessing

After obtaining the data, the data needs to be preprocessed to extract the data information. In this paper, the visualization method of financial data is selected, that is, the candlestick chart preprocesses the stock data. Use Matplotlib library [22] in Python to convert the historical data of time series into candlestick charts. The candlestick chart used in this study is shown in Figure 4, which is based on different markers for comparative experiments. In order to analyze the relationship between different forecast intervals and forecast accuracy, this article marks each candle chart based on different forecast intervals (close difference of 1, 20, 30, 60 and 90 trading days), The calculation method is as formula (1), when $close_{i+d} < close_i$, $target_i = 0$, when $close_{i+d} > close_i$, $target_i = 1$. In addition to the different time intervals, whether the volume information is included can also be used as a basis for distinguishing the two sets of data (see Figure 3 a) and Figure 3 b）). In addition, the candlestick chart (see Figure 3c) with the addition of the MACD indicator and the moving average and the GAF picture (see Figure 3 d)) were used as a comparison between the other two groups. At this point, there are three variables, namely prediction interval, picture type and classification method.The candlestick charts used in this article all contain 60 days of information [23].

$$target_i = sign(close_{i+d} - close_i) \qquad (1)$$

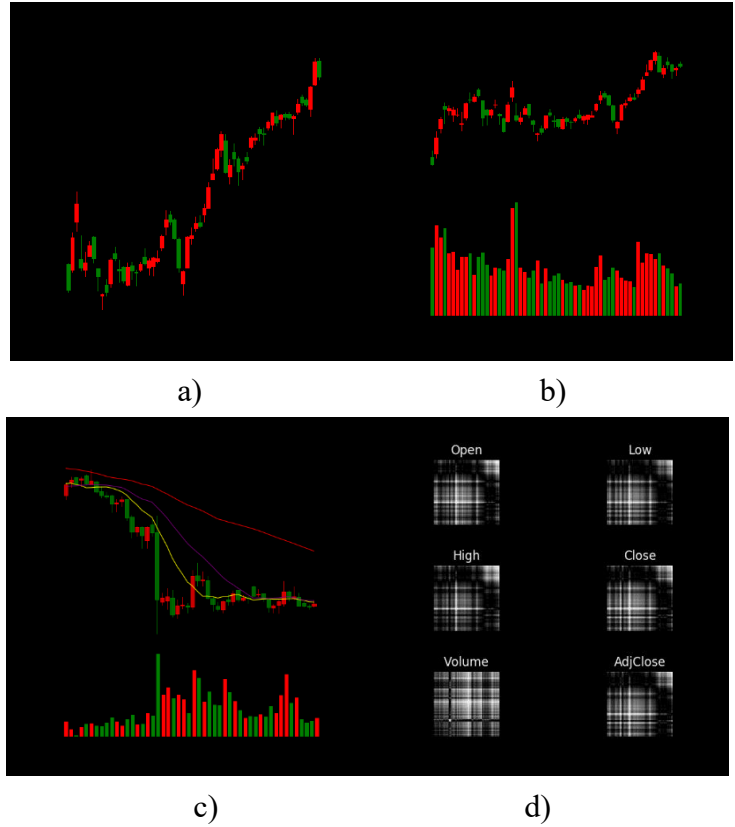

a)                      b)

c)                      d)

**Figure 4.** Four candles of different richness

## 4. Results

In this section, first calculate the prediction accuracy based on the obtained CNN model, and then analyze the impact of the three indicators of prediction interval, picture richness, and classification on the prediction accuracy. It can be concluded that a better and accurate data set type can be obtained for further research in the future.

4.1 Grade

There are some statistical methods for performance evaluation, which can evaluate the results of all classifiers by measuring sensitivity (true positive rate or recall rate), specificity (true negative rate), accuracy, and Matthew correlation coefficient (MCC). Generally, TP is true positive or correct recognition, FP is false positive or false recognition, TN is true negative or correct rejection, and FN is false negative or false rejection. The corresponding formula is as follows:

$$Sensitivity = \frac{TP}{TP+FN} \tag{2}$$

$$Specitivity = \frac{TN}{TN+FP} \tag{3}$$

$$Accuracy = \frac{TN + TP}{TP + FP + TN + FN} \qquad (4)$$

$$MCC = \frac{TP \times TN - FP \times FN}{\sqrt{(TP+FP)(TP+FN)(TN+FP)(TN+FN)}} \qquad (5)$$

4.2 Impact of Prediction Interval on Accuracy

From Figure 5a), the first type of graph with the least information richness, the prediction accuracy increases with the time interval. The accuracy at the interval of 1 is 0.489; the accuracy of 20-60 days increases to 5%; until the interval exceeds 90 days, the accuracy exceeds 0.6, and it is considered that this division method is not well suited for the task of describing the stock market forecast.

As can be seen from Figure 5 b), the second type of graph with volume information has an accuracy of 0.495 at the interval of 1 to 0.597 at the interval of 30, and further to 0.558 at the interval of 90. There is a maximum value , The maximum interval is 30 days.And the maximum accuracy within the test interval is lower than that of the first type.

It can be obtained from Figure 5 c) that as the graph with the highest degree of information integration in the first three categories of graphs, the lowest accuracy is 0.603 at 90-day intervals and the highest accuracy is 0.922 at 20-day.Compared with the previous two types of graphs, the highest accuracy improvements are 52.4% and 54.4%, respectively, which have application value.

Figure 5 d) is the GAF graph, and the time interval has little effect on it. The difference between the maximum (90-day interval) accuracy and the minimum (30-day interval) accuracy is only 0.068.

In summary, in the case of the third type of images with the highest image richness, CNN can extract enough feature values to train the model for the prediction, and the best prediction interval of the mark is about 20 days at this time. More accurate results can be obtained through subdivision comparison. This article only focus on rough intervals and does not pursue further comparerisons.

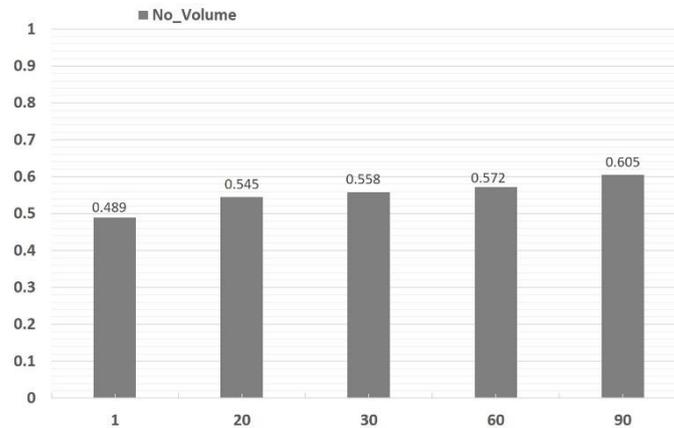

a) No Volume

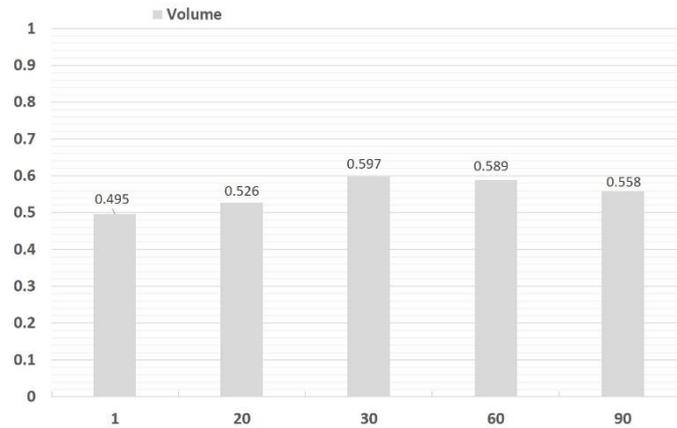

b) Volume

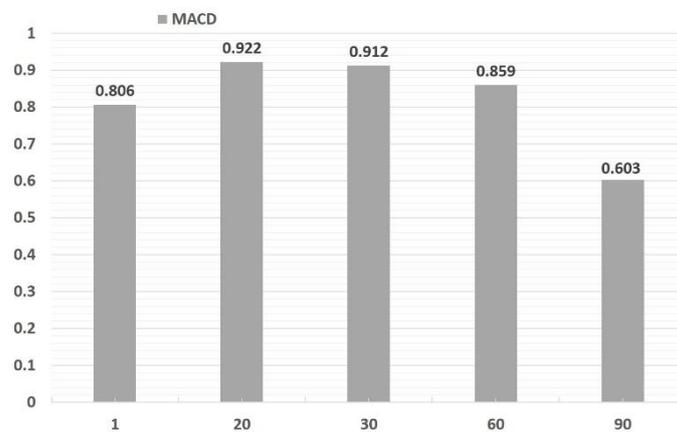

c) MACD

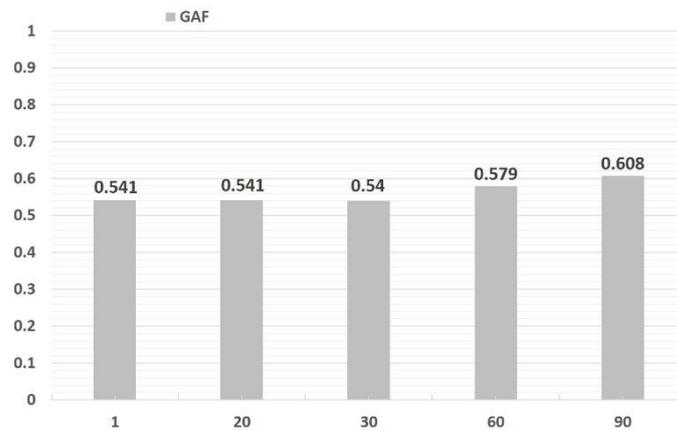

d) GAF

**Figure 5.** Accuracy comparison of four types of candle charts at different intervals

4.3 Impact of Image Richness on Accuracy

We can sort the four types of candlesticks based on the richness of information, that is, the third type>fourth type>second type>first type.

Based on the analysis in 4.2, we can clearly see that different from even the lowest prediction accuracy of the third category graph and the highest accuracy of the other three category graphs, is only 0.002-0.003.

The results are obvious. The third type of chart, that is, the stock chart containing the MACD indicator and three moving averages (M5, M10, M30), has a significantly higher accuracy than the other stock charts at each forecast interval.

Therefore, within the scope of the research in this article, the third type of graph is the best prediction result as a data set training. The comparison of the training accuracy of the four types of pictures is shown in Figure 6.

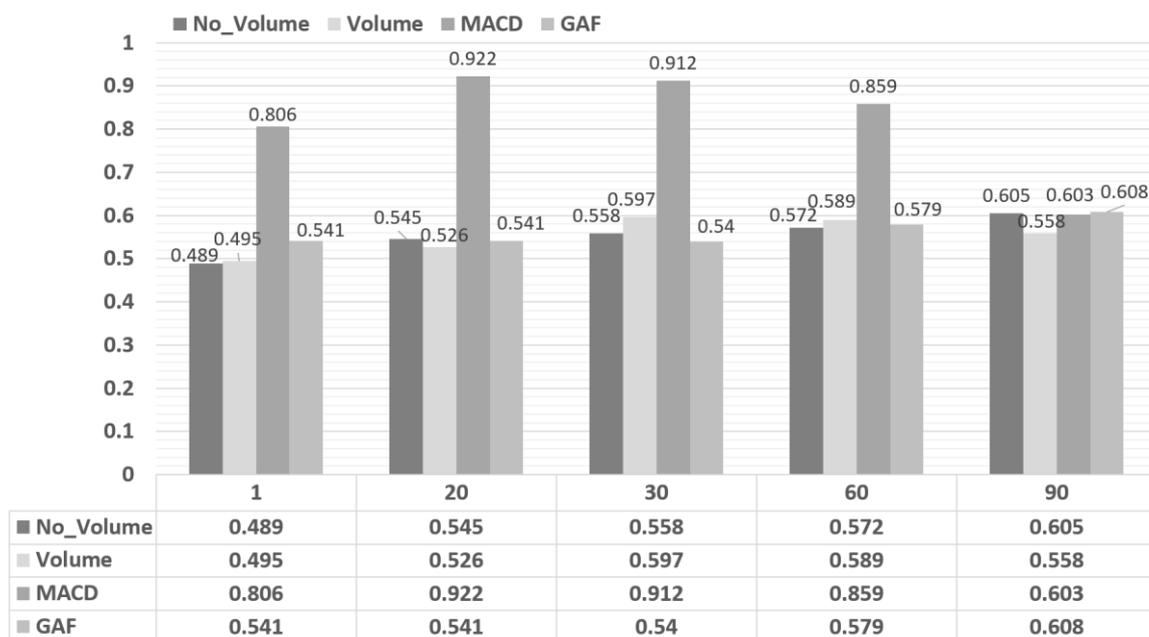

**Figure 6.** CNN model prediction accuracy

4.4 Impact of Classification on Accuracy

This article compares the classification methods based on each candle chart containing 60 days of data and using 90-day forecast interval tags. The results are shown in Figure 7.

It can be seen from the figure that the accuracy of random division is similar to that of automatic division, which is higher than that of time series division. However, the random division method can cause the time series of the training set and the prediction set to be mixed, which may cause the situation of predicting the intermediate result under the premise of the known result, so the random division method is not accurate. In addition, the automatic division method may have a greater impact on the prediction accuracy because of the similarity of the content of the data set. The higher the similarity of the stock data set, the lower the impact; the smaller the similarity, the lower the accuracy of the test.

We select The NASDAQ exchange, which is established in the 1970s, as one of the largest stock exchanges in the world. Many large international companies are listing on the NASDAQ, with huge market value, those companies have high similarityies in stock movement trends. The stocks of these companies generally fluctuate heavyly under the influence of policies, and the trend of ups and downs are similar and time-

sensitive.This is why the accuracy of automatic division in this article is so high. Of course, it is also possible that the stocks in the second half have continued to rise or fall over the past four years. In this case, the stock trend is very easy to judge. This is also the reason why the accuracy may reach 0.9.

Therefore, despite the lowest accuracy, this paper still chooses the time series partition method to partition the training set and the test set.

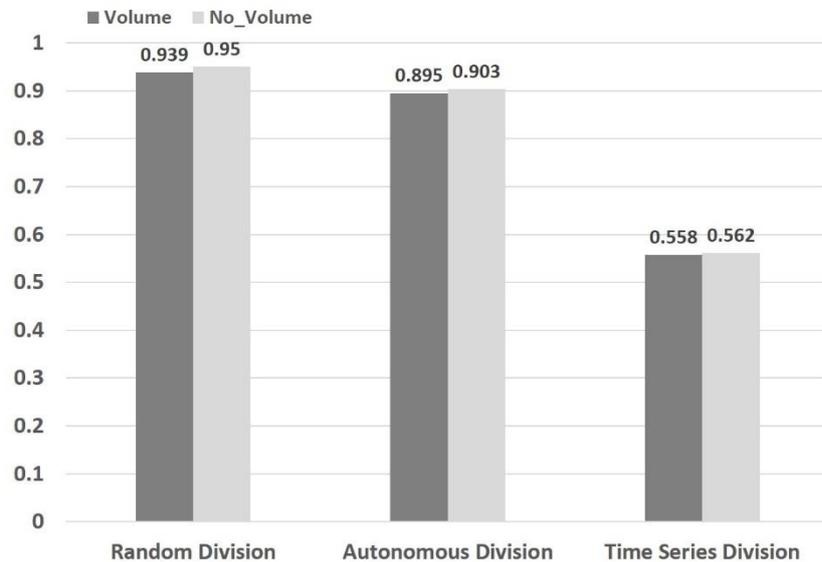

**Figure 7.** Comparison of the impact of classification methods on CNN prediction

4.5 Results Expansion

It can be seen from 4.3 that the accuracy of the stock chart with MACD information has obvious advantages over the other three categories. On this basis, the division of the picture area is changed. Place the MACD indicator, Volume information and K line at the positions shown in Figure 7 a) (place the Volume and MACD indicators in the lower part of the picture, and the K line position remains unchanged). When the remaining variables are unchanged, the data set is put into CNN training, and the obtained accuracy comparison chart is shown in Figure 7 b).It can be found from the figure that by changing the division area, the accuracy of the CNN model can be increased by 0.6%-2.2%, indicating that on the basis of sufficient richness, changing the division area can improve the prediction accuracy. The future we will continue to explore towards this direction.

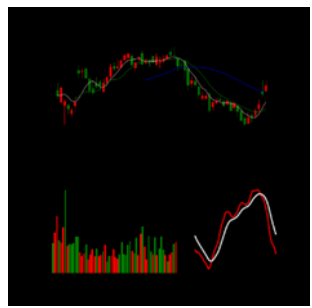

a) 对比图

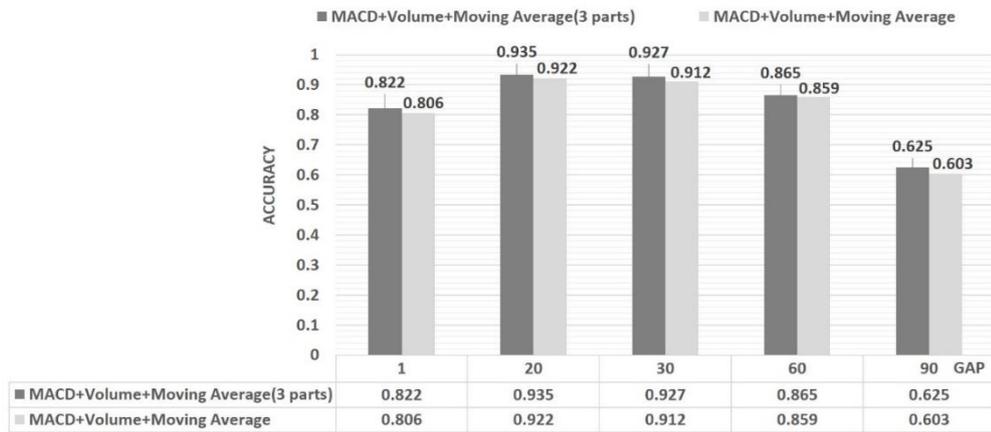

b) 准确度对比图
**Figure 7.** Expand pictures and data

## 5. Conclusion

This article collected and processed the transaction time series data of NDAQ 100 stocks from December 31, 2014 to December 31, 2019, fed into self-built stock market prediction model based on CNN, using statistical indicators to perform quantitative analize of the results. The conclusions are as follows:

1) Observed the prediction results of the CNN model, we could see that the stock chart with MACD indicators and three moving averages (M5, M10, M30) have an accuracy of 0.922 at a time interval of 20 days, which is the best combination of all experimental conditions.
2) For the three classification methods, random division will cause the time series of the training set and the prediction set to be mixed, and the automatic division method may cause a large error in prediction accuracy because of the similarity of the data. Therefore, the least accurate time series division method is selected.
3) From the expansion results, we can see that the training results of the data set after region division have improved. Future research will be conducted on the basis of this expansion.